\documentclass[manuscript,screen]{acmart}

\AtBeginDocument{%
  \providecommand\BibTeX{{%
    \normalfont B\kern-0.5em{\scshape i\kern-0.25em b}\kern-0.8em\TeX}}}

\setcopyright{acmcopyright}
\copyrightyear{2018}
\acmYear{2018}
\acmDOI{10.1145/1122445.1122456}

\acmJournal{JACM}
\acmVolume{37}
\acmNumber{4}
\acmArticle{111}
\acmMonth{8}

\begin{document}

\newcommand{\hashtag}[1]{{\textit{\##1}}}
\newcommand{\group}[1]{{\texttt{#1}}}
%%
%% The "title" command has an optional parameter,
%% allowing the author to define a "short title" to be used in page headers.
\title{How Twitter Data Sampling Biases U.S. Voter Behavior Characterizations}

%%
%% The "author" command and its associated commands are used to define
%% the authors and their affiliations.
%% Of note is the shared affiliation of the first two authors, and the
%% "authornote" and "authornotemark" commands
%% used to denote shared contribution to the research.

\author{Kai-Cheng Yang}
\email{yangkc@iu.edu}
\orcid{0000-0003-4627-9273}
\affiliation{%
  \institution{Observatory on Social Media, Indiana University}
  \city{Bloomington}
  \state{Indiana}
  \country{USA}
  \postcode{47408}
}
 
\author{Pik-Mai Hui}
\affiliation{%
  \institution{Observatory on Social Media, Indiana University}
  \city{Bloomington}
  \state{Indiana}
  \country{USA}
  \postcode{47408}
}

\author{Filippo Menczer}
\affiliation{%
  \institution{Observatory on Social Media, Indiana University}
  \city{Bloomington}
  \state{Indiana}
  \country{USA}
  \postcode{47408}
}

%%
%% By default, the full list of authors will be used in the page
%% headers. Often, this list is too long, and will overlap
%% other information printed in the page headers. This command allows
%% the author to define a more concise list
%% of authors' names for this purpose.

\renewcommand{\shortauthors}{Yang et al.}

%%
%% The abstract is a short summary of the work to be presented in the
%% article.
\begin{abstract}
Online social media are key platforms for the public to discuss political issues.
As a result, researchers have used data from these platforms to analyze public opinions and forecast election results.
Recent studies reveal the existence of inauthentic actors such as malicious social bots and trolls, suggesting that not every message is a genuine expression from a legitimate user. 
However, the prevalence of inauthentic activities in social data streams is still unclear, making it difficult to gauge biases of analyses based on such data. 
In this paper, we aim to close this gap using Twitter data from the 2018 U.S. midterm elections. 
Hyperactive accounts are over-represented in volume samples. 
We compare their characteristics with those of randomly sampled accounts and self-identified voters using a fast and low-cost heuristic.
We show that hyperactive accounts are more likely to exhibit various suspicious behaviors and share low-credibility information compared to likely voters.
Random accounts are more similar to likely voters, although they have slightly higher chances to display suspicious behaviors.
Our work provides insights into biased voter characterizations when using online observations, underlining the importance of accounting for inauthentic actors in studies of political issues based on social media data.
\end{abstract}

%%
%% The code below is generated by the tool at http://dl.acm.org/ccs.cfm.
%% Please copy and paste the code instead of the example below.
%%


%%
%% Keywords. The author(s) should pick words that accurately describe
%% the work being presented. Separate the keywords with commas.

%%
%% This command processes the author and affiliation and title
%% information and builds the first part of the formatted document.
\maketitle

\section{Introduction}

In recent years, social media have been serving as important platforms for news dissemination and public discussion of politics.
With more citizens consuming information and actively expressing their political opinions online, political figures, governments, and agencies have started to adopt social media to reach out to the public and amplify their influence \cite{jungherr2016twitter}.
These activities generate huge amounts of valuable data, making it possible to study voter behavior \cite{small2011hashtag,jungherr2014logic,ausserhofer2013national} and estimate public opinions around political issues \cite{diakopoulos2010characterizing,wang2012system} through computational approaches.
Some researchers even attempt to predict election outcomes \cite{tumasjan2010predicting,digrazia2013more,burnap2016140}, although the reliability of such predictions has been questioned \cite{gayo2011limits}.

\begin{figure}
    \centering
    \includegraphics[width=0.8\columnwidth]{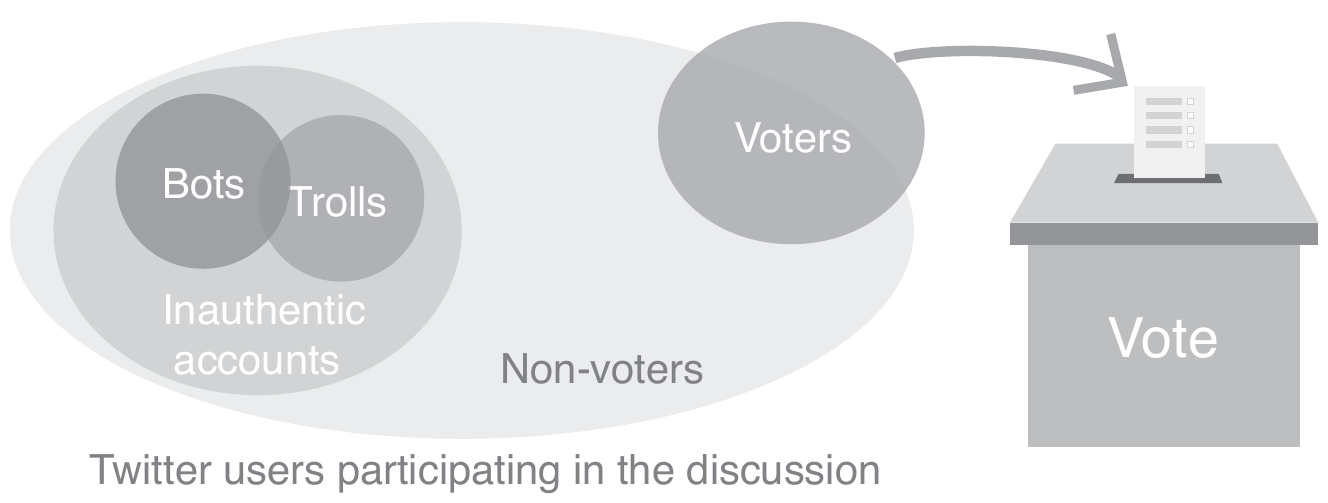}
    \caption{
    The composition of different sets of accounts participating in  election-related discussion on Twitter.
    The voters determine real-world outcomes, while the online data comes from a population that includes non-voters and inauthentic accounts. The sizes of the sets in the diagram do not correspond to the actual numbers of accounts in each group.
    }
    \label{fig:venn_diagram}
\end{figure}

However, raw social media data might be seriously biased and even lead to false signals.
To demonstrate, we use Twitter as an example of social media platforms given its popularity among U.S. politically active citizens. % citation?
Fig.~\ref{fig:venn_diagram} illustrates the composition of the different types of accounts that participate in the online discussion about elections.
Ideally, we want to focus on the data generated by real-world voters since they determine the election outcomes.
But the data stream originates from a population that does not contain all voters and that contains non-voters and inauthentic accounts.
The latter group includes entities like malicious social bots and foreign trolls that have been shown to manipulate public opinion regarding elections \cite{bessi2016social,deb2019perils,stella2018bots,ferrara2017disinformation,badawy2018analyzing,shao2018spread,zannettou2019disinformation}.
Relying on content generated by such accounts is clearly problematic.

Although bots and trolls have been studied extensively, we conjecture that they do not account for all inauthentic behavior on Twitter.
To understand potentially biased characterizations of U.S. voters based on observations of Twitter data, it is important to quantify the prevalence of different questionable behaviors.
To this end, here we create two different samples of accounts during the 2018 U.S. midterm elections. First, naive sampling based on tweet volume leads to over-representation of hyperactive accounts. Second, random resampling of unique accounts removes this bias. 
We also propose an efficient method to extract  self-identified voters.
We systematically examine and compare the characteristics of accounts selected according to these different sampling methods.  
We analyze account profiles, activity in the discussion, general sentiment expression, political alignment, sharing of low-credibility information, and various suspicious behaviors.

We find that hyperactive accounts are more likely to exhibit various suspicious behaviors and share low-credibility information compared to likely voters.
These findings further our understanding of biases in characterizations of U.S. voters based on Twitter data, providing useful guidance toward political studies on social media. 
For journalists and the public at large, it reveals important insights about the accounts encountered online everyday. 

\section{Related Work}

\subsection{Politics-related Social Media Studies}

Many recent computational social science and political science studies are dedicated to understanding online behaviors of voters in political discussions.
To name a few, \citet{small2011hashtag} studies the usage of political hashtags.
\citet{jungherr2014logic} investigates the logic of political coverage on Twitter by comparing it with traditional media.
And \citet{ausserhofer2013national} study interrelations between the users.

Another line of work seeks to predict election outcomes using indicators estimated from social media data.
In the context of a German election, \citet{tumasjan2010predicting} show that the number of tweets can reflect the election result.
A study by \citet{digrazia2013more} similarly finds a correlation between the votes and candidate name mentions on Twitter in the 2010 U.S. midterm election, but also accounts for other predictors.
Considering the same election, the model proposed by \citet{livne2011party} uses variables derived from Twitter content and network features and achieves decent prediction performance.
Other studies depend on the sentiment of the tweets. 
For example, \citet{o2010tweets} find that tweet sentiment is correlated with presidential job approval and consumer confidence polls.
\citet{ceron2014every} demonstrate the usefulness of sentiment analysis in multiple elections.
\citet{burnap2016140} show similar results in the 2015 U.K. general election.
The reliability of electoral predictions based on social media data have also been questioned \cite{gayo2011limits}.

\subsection{Questionable Actors and Behaviors}

The politics-related social media studies reviewed above usually assume that social media data comes from real users and barely mention inauthentic accounts.
This is not surprising, as studies of inauthentic actors on social media have only started to emerge in recent years.

Social bots are social media accounts controlled completely or in part by algorithms \cite{stocking2018social,ferrara2016rise}.
Their automated nature allows a bot master to easily generate large amounts of content or create false appearance of popularity. 
Recent reports show that malicious bots were involved in recent elections in Europe and U.S. \cite{bessi2016social,deb2019perils,stella2018bots,ferrara2017disinformation}.
Close analyses suggest that bots tend to have suspicious profile information, recent creation dates, large numbers of tweets, and limited numbers of original posts \cite{ferrara2016rise}. 
Questionable actions of malicious bots include astroturfing \cite{Truthy_icwsm2011politics}, spreading misinformation \cite{shao2018spread}, and using inflammatory language to irritate others \cite{stella2018bots}.

In this paper we use the term ``trolls'' to refer to fake accounts controlled by state-sponsored agencies. 
Studies suggest that these accounts have interfered with the 2016 U.S. presidential election \cite{badawy2018analyzing,zannettou2019let,zannettou2019disinformation}. 
The characteristics of the trolls can be different from those of random Twitter accounts: they tend to use deceptive language \cite{addawood2019linguistic}, their posts come from different Twitter clients, and their creation dates are concentrated \cite{zannettou2019disinformation}.

Note that the bot and troll behaviors mentioned above may only represent a subset of inauthentic behaviors online; other types of suspicious behaviors may remain unnoticed. As a result, our understanding of the prevalence of questionable behaviors among typical Twitter users is limited. 

\subsection{Voter Identification}

On social media, voters are often identified by conducting surveys.
\citet{mellon2017twitter} use a nationally representative sample in the U.K. and find that Twitter and Facebook users tend to be younger, well-educated, more liberal, and to pay more attention to politics.
The Pew Research Center has a series of studies in which they first survey a group of U.S. adults from a nationally representative panel for their demographics and political ideology.
Then they link the survey data to Twitter handles shared by the participants \cite{wojcik2019sizing,hughes2019national}. Among other things, they find that a small portion of users generate most of the political content on Twitter, politics is not the main topic of conversation for voters. 

Another approach starts from social media handles and then attempts to establish links to voters.
\citet{barbera2016less} propose a method to match geolocated tweets with voting registration records to obtain the demographics of the people behind the Twitter handles.
\citet{grinberg2019fake} match Twitter profiles against voter registration information, finding that a small portion of U.S. voters account for the majority of fake news exposure and sharing during the 2016 presidential election.

The approaches introduced above can be powerful in connecting online behaviors and demographics of voters. 
However, high quality data sources like nationally representative panels, voting registration records, and large-scale surveys are expensive and out of reach for many researchers. 
Moreover, these approaches can hardly be automated.
In their recent work, \citet{deb2019perils} collect tweets related to the 2018 U.S. midterm elections and treat accounts tweeting the hashtag \hashtag{ivoted} on election day as actual voters.
This method can be applied to large-scale data efficiently and thus makes rapid or even real-time analysis possible, although we show here that it introduces some bias.

\section{Data}

\subsection{Tweet Collection}

In this study, we focus on Twitter data around the 2018 U.S. midterm elections.
To create a dataset containing most of the election-related content, we collect tweets that include any hashtag in a query through Twitter's free filtering API.
The query includes a list of 143 election-related hashtags from a previous study \cite{yang2019bot}.
The list was initialized with several well-known hashtags like \hashtag{maga}, \hashtag{bluewave}, and \hashtag{2018midterms}.
A snowball sampling method based on co-occurrence was then applied to expand the list \cite{Truthy_icwsm2011politics}.
Hashtags used for each state's Senate election (e.g., \hashtag{AZsen} and \hashtag{TXsen}) were also added to the list.
We then manually check the list and removed irrelevant hashtags.
This method is designed to capture most of the election-related tweets with high precision; although the list was not updated during data collection, newly emerging political hashtags likely co-occurred with those already on the list. 
The collection started on Oct 5, 2018 and continued until the end of 2018, resulting in a dataset with about 60M tweets by 3.5M unique users.

\subsection{Naive Accounts Sampling}

Our goal is to understand the characteristics of typical Twitter users and compare them with identified voters.
For a meaningful comparison, we use the collected dataset as the sampling pool, i.e., we focus on accounts who discussed the 2018 U.S. midterm elections. 
We deploy two different methods to sample typical accounts.
With an unweighted sampling method, we simply select unique accounts in our dataset randomly, hoping to get a representative sample.
We call the group of accounts obtained with this sampling method \group{ByAccount}.

Sampling users from the tweets in our dataset favors active accounts: the chance of being selected is proportional to the number of tweets each account has in our dataset. 
We call the group of accounts sampled by this method \group{ByTweet}.
In the following analyses, we will show that \group{ByTweet} accounts are hyperactive and generate disproportionately more content.
As a result, Twitter users are more likely to be influenced by these accounts directly or indirectly, and content analysis based on the Twitter stream could be biased accordingly.
The \group{ByAccount} and \group{ByTweet} groups have the size as the \group{Voter} group, described next.

\section{Identification of Voters}

In this section we present high-precision methods to identify likely voters on Twitter using the data collected.
These accounts are used as a reference group to describe authentic behaviors in the online discussion regarding U.S. politics.

The basic idea is to use the messages and pictures posted by users on election day.
We follow the approach of \citet{deb2019perils} and use the hashtag \hashtag{ivoted} as a first filter.
The hashtag was promoted by various social media platforms to encourage people to vote in the midterm elections, and it became quite popular for users to tweet about it after voting on election day (Nov 6, 2018).
However, considering all accounts tweeting the \hashtag{ivoted} hashtag as voters may result in false positives since inauthentic accounts can also post the hashtag.
In fact, Deb et al.\ exclude about 20\% of the accounts that tweeted \hashtag{ivoted} in their analysis, based on high Botometer scores \cite{yang2019arming}.

Instead of attempting to exclude all problematic accounts from those tweeting \hashtag{ivoted}, which is itself prone to false positive and false negative errors, we aim to include only those who are very likely to be real voters. This approach yields voter accounts with high precision at the expense of recall, but this is acceptable for our purpose of characterizing likely voters. 
We notice that many voters shared pictures of them wearing ``I Voted'' stickers or buttons after voting (see Fig.~\ref{fig:vote_pics}). 
Based on this observation, we apply a second filter that verifies whether the \hashtag{ivoted} hashtag is accompanied by a picture showing a real person having voted.
%Although this procedure may lead to some false negatives, the confidence level of the identification is very high.

\subsection{Manual Annotation}

\begin{figure}
    \centering
    \includegraphics[width=0.8\columnwidth]{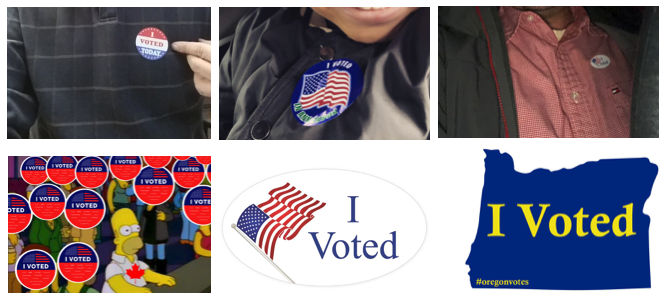}
    \caption{
    Examples of images providing positive (top) and negative (bottom) cases in our manual annotation of voters. 
    Images are cropped for privacy consideration.
    }
    \label{fig:vote_pics}
\end{figure}

To identify the voters in our dataset, we first extract all the tweets posted on election day that include both the \hashtag{ivoted} hashtag and images.
We download all the pictures for manual annotation.
A single criterion is used in the annotation: whether we can reach the conclusion that a real person just voted based on the picture.
The annotations based on this criterion turn out to be quite straightforward. 
In most of the positive cases, the picture clearly shows a person with an ``I Voted'' sticker or button on their cloth or body.
There are also some borderline cases where the pictures show voting-related memes or computer-generated ``I Voted'' stickers.  
Although the corresponding users might have voted, we still consider these cases negative because sharing these pictures can be faked easily. 
Fig.~\ref{fig:vote_pics} shows examples of both positive and negative cases.

Although \hashtag{ivoted} is not on the list of hashtags used for collecting the tweets, it often co-occurs with other hashtags in the query, so we end up with 40,078 positive tweets. 
Among these, 13,084 have pictures attached.
Through annotation and data cleaning, we identify 8,092 voters.
We call this group of accounts \group{Voter}.
For comparison, we also sample an equal number of accounts that tweeted \hashtag{ivoted} with no pictures and call this group \group{Ivoted}.

The free Twitter API is sufficient for collecting the data.
Manual annotation is relatively time-consuming: it took about 30 hours for the first author to go through all the images. 
But the task is relatively straightforward, so no special training is needed.
Moreover, implementing this method only requires basic programming skills and a Twitter developer account.

\subsection{Automated Annotation}

Since the human annotation is the bottleneck of the method described above, we also implemented an automated annotation process.
The key challenge is to translate the annotation criterion into a format that machines can understand.
Since most of the positive cases have the text ``vote'' in the picture, we operationalize the annotation criterion as identifying whether the picture contains the string ``vot'' (to match ``vote,'' ``voted,'' and ``voting'').

Traditional optical character recognition software\footnote{\url{https://opensource.google/projects/tesseract}} did not provide acceptable performance.
Therefore we resorted to an off-the-shelf deep learning model provided by the Google Cloud service.\footnote{\url{https://cloud.google.com/}} 
We send all images to the the text detection API endpoint to extract text strings and check whether ``vot'' is in the response.

In our experiment, the automated approach annotates 46.5\% of the pictures as positive.
By treating the manual annotations as ground truth, we measure the performance of the automated approach: it achieves 0.81 in precision and 0.59 in recall.
Close inspection of misclassified data reveals that most false positive cases are just like the ones in the bottom row of Fig.~\ref{fig:vote_pics}, as expected.
Typical false negative cases, on the other hand, are mostly due to limitations of the API in handling blurred, distorted, and rotated pictures.
For example, the API fails to detect the text in the upper right image in Fig.~\ref{fig:vote_pics}.
We collect the accounts labeled as voters by the automated method and call the group \group{Voter\_auto}.

With the help of Google Cloud or similar services, the whole voter identification process can be automated. The method is therefore fast and cheap; in our experiment, it takes less than 15 minutes and about \$20 to annotate all the pictures.
In summary, it is possible to setup an infrastructure to automatically identify likely voters on Twitter in real time at a very low price, if low recall is acceptable. 
With the rapid development of deep learning technologies, we expect to see an increase in performance and a decrease in price over time.

\section{Characterization of Account Behaviors}

In this section, we compare various characteristics of accounts in \group{Voter}, \group{Voter\_auto}, \group{Ivoted}, \group{ByTweet}, and \group{ByAccount} groups.

\subsection{Account Profiles}

\begin{figure}
    \centering
    \includegraphics[width=\columnwidth]{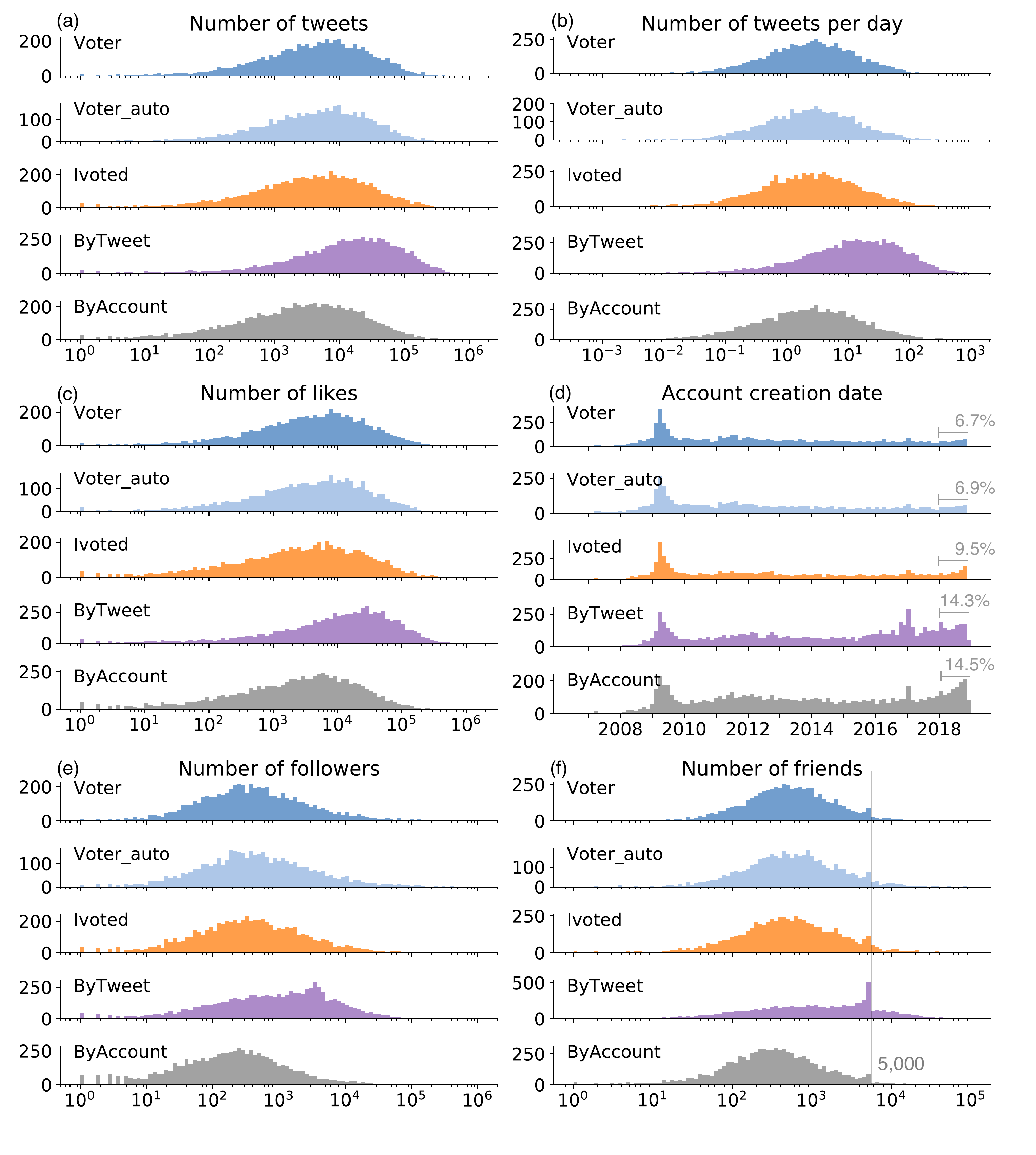}
    \caption{
    Distributions of (a)~number of tweets; (b)~number of tweets per day; (c)~number of likes; (d)~account creation date; (e)~number of followers; and (f)~number of friends for different groups.
    The annotations in (d) represent the percentage of accounts that were created in 2018 for each group.
    The line in (f) indicates 5,000 friends.
    }
    \label{fig:basic_stats}
\end{figure}

We systematically compare the profiles of the accounts in different groups and show selected results in Fig.~\ref{fig:basic_stats}.
The information can be accessed directly from the user object embedded in each tweet, which reflects the status of the account at the time the tweet was collected.
Since these user objects may change over time, we use the earliest version of each account in the collection for our comparison. 

We show the distributions of total number of tweets in Fig.~\ref{fig:basic_stats}(a).
It appears that \group{ByTweet} accounts generate more tweets than other groups across their life time.
Since the total number of tweets depends on the account age, we also show the distributions of number of tweets per day in Fig.~\ref{fig:basic_stats}(b).
The age of an account is defined as the time difference between the date of the data collection and the account creation date.
Again, we can see that \group{ByTweet} accounts tweet faster than other groups.
Over a half of the \group{ByTweet} accounts post more than 10 tweets per day; some post more than 100 tweets per day.
The distributions of the two measures for other account groups are similar.

We also examine the number of likes and plot the distributions in Fig.~\ref{fig:basic_stats}(c).
Again, \group{ByTweet} accounts have more likes than other groups, which have identical distributions.
The distributions of the number of likes per day (not shown) suggest that \group{ByTweet} accounts produce likes faster as well.

The distributions of account creation time are presented in Fig.~\ref{fig:basic_stats}(d).
Across all groups, we observe a peak around 2009, which is likely caused by the fast growth of Twitter during that year.
\group{ByTweet} and \group{ByAccount} groups have more accounts created in 2018 ---right before the midterm elections--- than other years.
We annotate the percentage for each group in the figure to confirm this observation.

Fig.~\ref{fig:basic_stats}(e) and (f) show the distributions of the numbers of followers and friends, respectively.
The number of friends indicates how many other accounts each account is following.
We find that \group{ByTweet} accounts tend to have more followers and friends than other groups.
Interestingly, \group{ByTweet} has more accounts with about 5,000 friends.
This is due to an anti-abuse limitation of Twitter, which states that each account can follow up to 5,000 friends unless they have more followers.\footnote{\url{https://help.twitter.com/en/using-twitter/twitter-follow-limit}}
The pattern here suggests that accounts in \group{ByTweet} are eagerly expanding their friend lists, until hitting the limit.
\group{ByAccount} accounts tend to have fewer followers and friends than the voter groups, although the difference is mild.

Through the analyses of the account profiles, we find that \group{ByTweet} accounts are not only hyperactive in terms of tweeting, as expected given the sampling method; they are also hyperactive in liking and building social networks, and they tend to be more recent. 

\subsection{Tweeting Behaviors during Elections}

\begin{figure}
    \centering
    \includegraphics[width=\columnwidth]{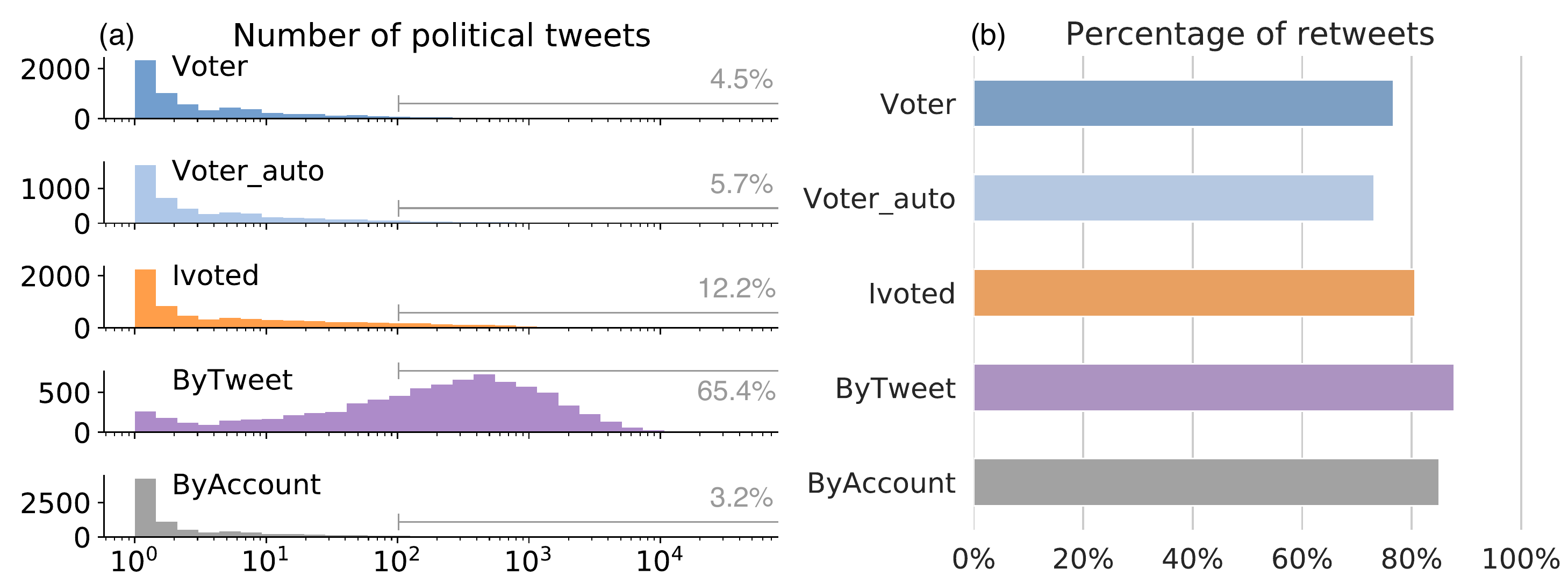}
    \caption{
    (a)~Distributions of number of political tweets for different groups.
    (b)~Percentage of retweets among the tweets generated by different groups.
    Annotations show percentages of accounts with over 100 political tweets.
    }
    \label{fig:participation}
\end{figure}

We examine the level of participation in the online discussion about the midterm elections for different groups. 
This is measured by the number of tweets each account has in our dataset (see the distributions in Fig.~\ref{fig:participation}(a)).
The three voter groups have distribution similar to \group{ByAccount}, with accounts typically having fewer than a hundred political tweets.
Due to the sampling method, it is expected that \group{ByTweet} accounts are more active in the discussion.
The discrepancy in political activity is huge: over 65\% of the \group{ByTweet} accounts have more than a hundred political tweets, compared 12\% or less in other groups.

These distributions suggest that the majority of the political tweets are generated by a small group of hyperactive users, which is in line with previous findings \cite{hughes2019national}.
This is concerning when one considers that volume of tweets is used to determine trending topics and viral content. Hyperactive accounts can flood the public data stream with content, drowning the voice of the majority of users and manipulating engagement metrics and mechanisms that rely on these metrics, such as feed ranking and trending algorithms. 

In addition to the absolute number of political tweets, we also examine the retweeting behaviors of different groups.
On Twitter, one can amplify messages from other accounts by retweeting or quoting them.
Quoting is relatively rare compared to retweeting, so we merge the two to simplify our analysis.
For the political tweets generated by each group, we calculate the percentage of retweets and show the results in Fig.~\ref{fig:participation}(b).
The majority of the tweets in each group are retweets, with \group{ByTweet} having a particularly high percentage.

\begin{figure}
    \centering
    \includegraphics[width=\columnwidth]{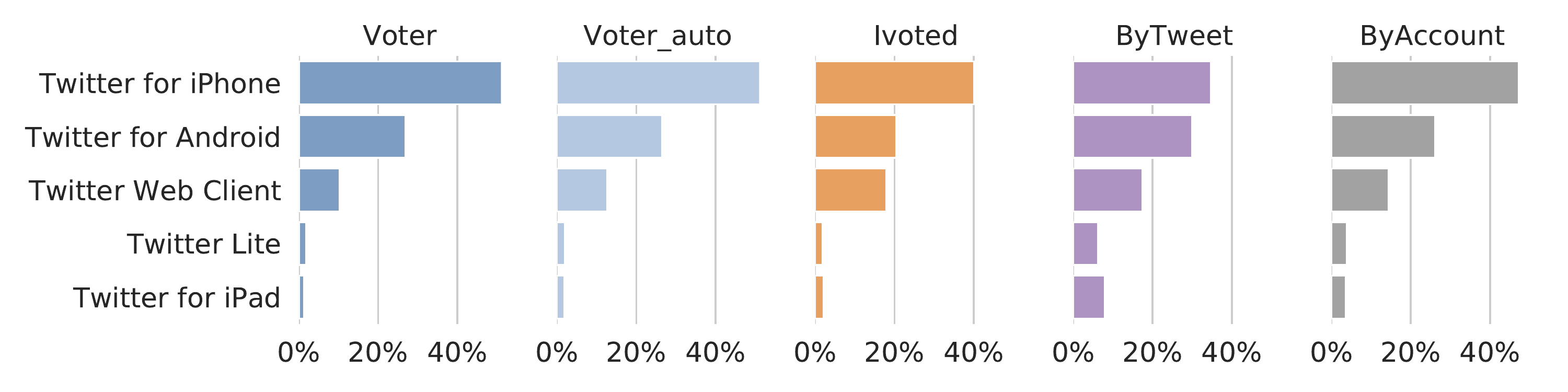}
    \caption{
    Share of Twitter clients for different groups.
    Only the five most popular clients are shown.
    }
    \label{fig:clients}
\end{figure}

Fig.~\ref{fig:clients} plots the share of Twitter clients in different groups.
The client information can be accessed from the tweet objects.
Since some accounts post tweets using multiple clients, we first obtain the most frequent client for each account and then count the clients within each group.
The major observation is the decreased share of iPhone clients and increased share of web clients in \group{ByTweet} and \group{Ivoted} when compared with other groups.
Similar patterns have been reported for trolls \cite{zannettou2019disinformation}.

\subsection{Sentiment Analysis}

\begin{figure}
    \centering
    \includegraphics[width=0.6\columnwidth]{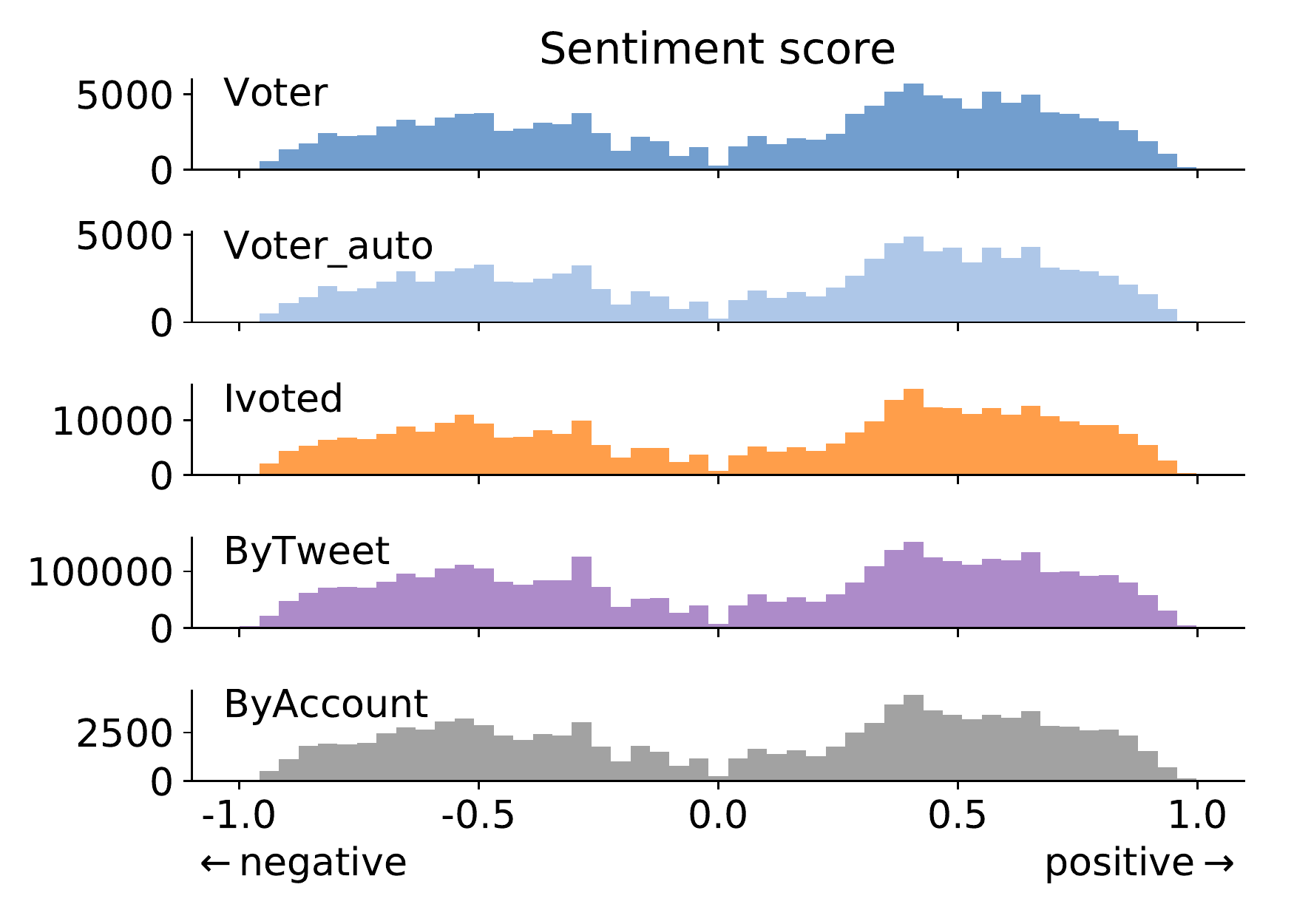}
    \caption{
    Tweet-level sentiment score distributions for different groups.
    Positive scores mean positive sentiment and vice versa.
    Sentiment-neutral tweets are excluded from this figure.
    }
    \label{fig:sentiment}
\end{figure}

Sentiment analysis is commonly adopted in mining social media for public opinions on political matters \cite{wang2012system,ramteke2016election,burnap2016140}.
Here we aim to compare the sentiment expressions of different groups.
We use VADER, an efficient lexicon- and rule-based sentiment analysis tool that works well on Twitter data \cite{hutto2014vader}.
In VADER's lexicon dictionary, each word and symbol is assigned a sentiment score through crowdsourcing. 
When applied to a tweet, VADER estimates the negative, positive, and neutral expressions to generate an overall sentiment score.

We show the tweet-level sentiment score distributions for different groups in Fig.~\ref{fig:sentiment}.
A score of zero is assigned to a tweet if it has no sentiment-expressing words.
The percentage of neutral tweets is consistent across the groups (about 30\%), so we exclude them from Fig.~\ref{fig:sentiment} to emphasize the patterns of positive and negative expressions.
We find that the general sentiment is consistent across groups.

\subsection{Political Alignment}

Since we are interested in behaviors regarding the U.S. elections, it is important to estimate the political alignment of accounts.
Following previous studies, we rely on the content produced by each account for this analysis.
We apply two independent methods, one using  hashtags and one using URLs (the links in the tweets) \cite{chen2020neutral}.
In both methods, a political alignment score is assigned to each entity (hashtag or URL) in the tweets.
The alignment score of each tweet is then obtained by averaging the associated entity scores, and the average score across an account's tweets is used as the alignment score of that account.
Next, we briefly introduce how each method assigns political alignment scores to the entities.

Hashtags are widely adopted by Twitter users to label topics and political alignment for their tweets.
Therefore it is common to use hashtags to infer user alignment  \cite{Truthy_icwsm2011politics,cota2019quantifying}.
Instead of manually labeling hashtags, we apply a recently developed approach to obtain continuous alignment scores in a semi-supervised fashion \cite{chen2020neutral}.
We first deploy the \textit{word2vec} algorithm \cite{mikolov2013efficient,mikolov2013distributed} to infer vectorized representations of the hashtags based on their co-occurrence.
Since the vectors encode the semantic relations between the entities, the axis between a pair of carefully selected hashtags can represent the political spectrum \cite{an2018semaxis,kozlowski2019geometry}.
The projected position of a hashtag on this axis reflects its alignment. 
In this study, we use \texttt{\#voteblue} and \texttt{\#votered} as the two anchors of the political alignment axis.
We then calculate the relative positions of the remaining hashtags on the axis to produce alignment scores in $[-1, 1]$ (negative values mean left-leaning).

Twitter users also commonly use links to share news and information from other platforms. 
Many websites, especially news outlets, have known political alignment, making it possible to infer an account's political score based on shared links. 
We adopt a list of 500 news sources with political alignment scores in $[-1, 1]$ (negative values mean left-leaning) from previous work \cite{bakshy2015replication}.
When processing the tweets, we extract the URLs, expand the shortened ones, obtain the domains, and then assign them scores from the list.
Entities and tweets with no score assigned are excluded in the evaluation.

\begin{figure}
    \centering
    \includegraphics[width=\columnwidth]{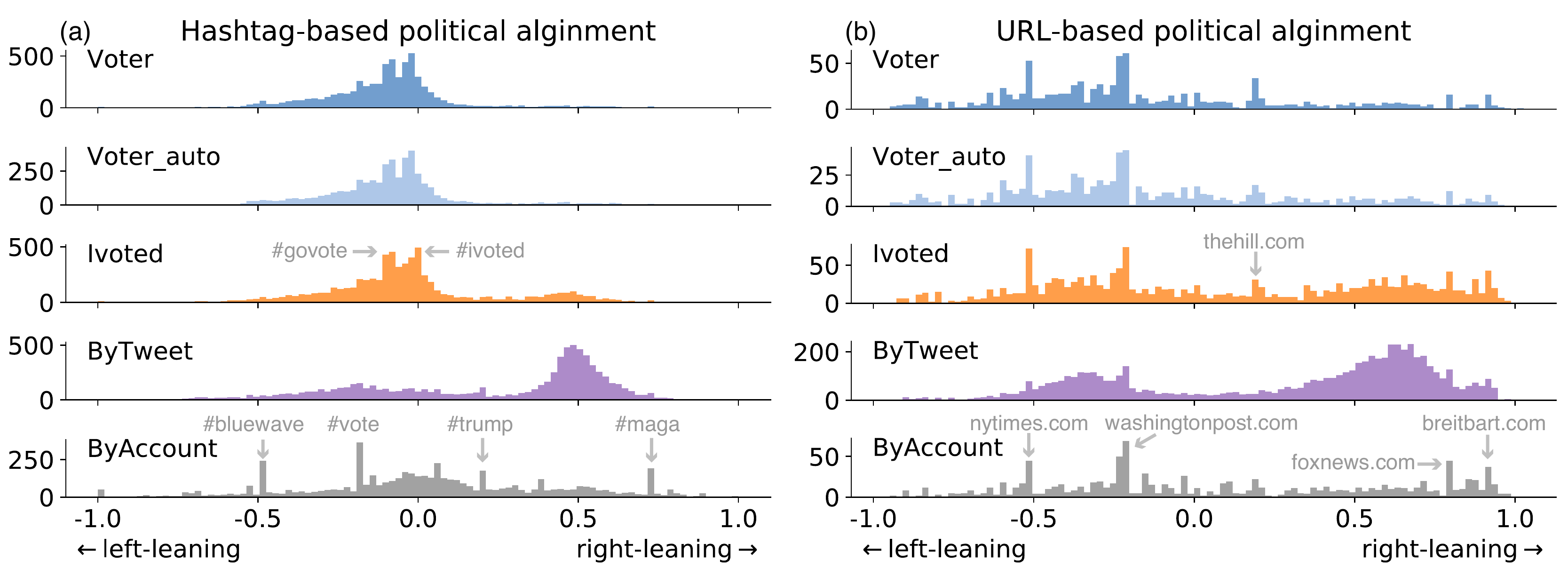}
    \caption{
    Distributions of political alignment scores for different groups using the (a)~hashtag-based and (b)~URL-based method.
    Peaks are due to popular entities, some of which are annotated.
    }
    \label{fig:political_alignment}
\end{figure}

We show the distributions of political alignment scores estimated using the two methods in Fig.~\ref{fig:political_alignment}(a,b). 
The two methods generate quantitatively different results but qualitatively similar patterns.
They suggest that voter accounts tend to be left-leaning, in agreement with previous surveys \cite{mellon2017twitter,wojcik2019sizing}.
\group{Ivoted} exhibits a similar distribution with \group{Voter} except that it has more right-leaning accounts.
The \group{ByAccount} group presents a relative diverse and symmetrical distribution across the spectrum.
Most surprisingly, accounts in the \group{ByTweet} group show a bimodal distribution with more accounts on the conservative side. This is inconsistent with the other groups, suggesting that the activity-weighted sampling bias completely distorts the political representation of Twitter accounts. 

\subsection{Automated Activities}

As discussed in the introduction, various inauthentic accounts might participate in the online discussion about the elections.
Therefore it is important to estimate the prevalence of these accounts in different groups.
Since social bots are the only type whose prevalence can be statistically estimated using off-the-shelf detection tools, we focus on  automated accounts in this subsection.

\begin{figure}
    \centering
    \includegraphics[width=0.6\columnwidth]{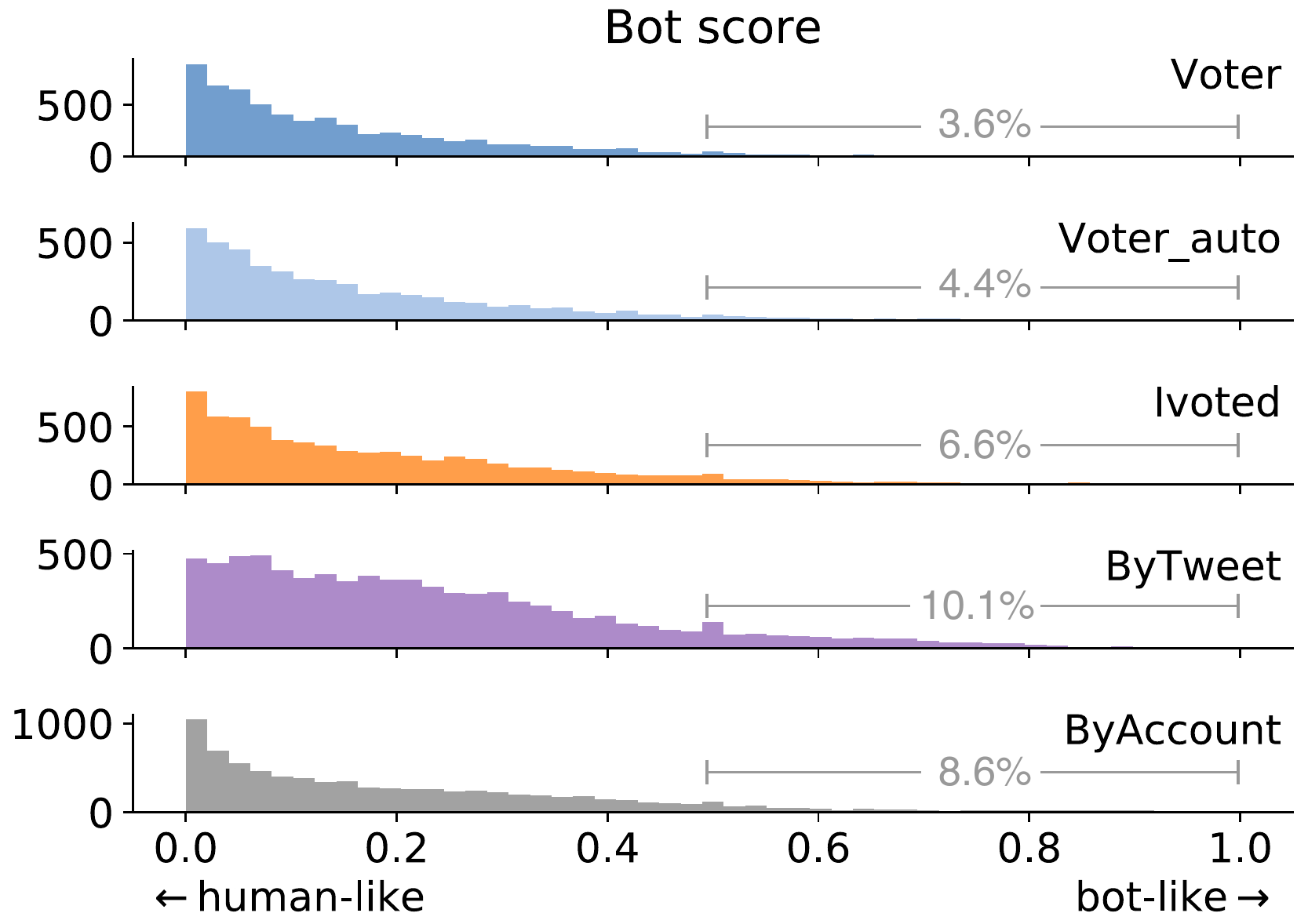}
    \caption{
    Bot score distributions of different account groups.
    We annotate the percentage of accounts having bot score above 0.5.
    }
    \label{fig:botscore_dist}
\end{figure}

We adopt the bot detection model proposed by \citet{yang2019scalable}.
By strategically selecting a subset of the training dataset, the model achieves high accuracy in cross-validation as well as cross-domain tests.
The classifier produces a bot score between zero and one for each account, where larger scores mean more bot-like. 
The bot score distributions for different groups are shown in Fig.~\ref{fig:botscore_dist}.
We can see that the \group{Voter} group has the lowest bot scores, while accounts in \group{ByTweet} tend to be more bot-like. 

\begin{figure}[t]
    \centering
    \includegraphics[width=\columnwidth]{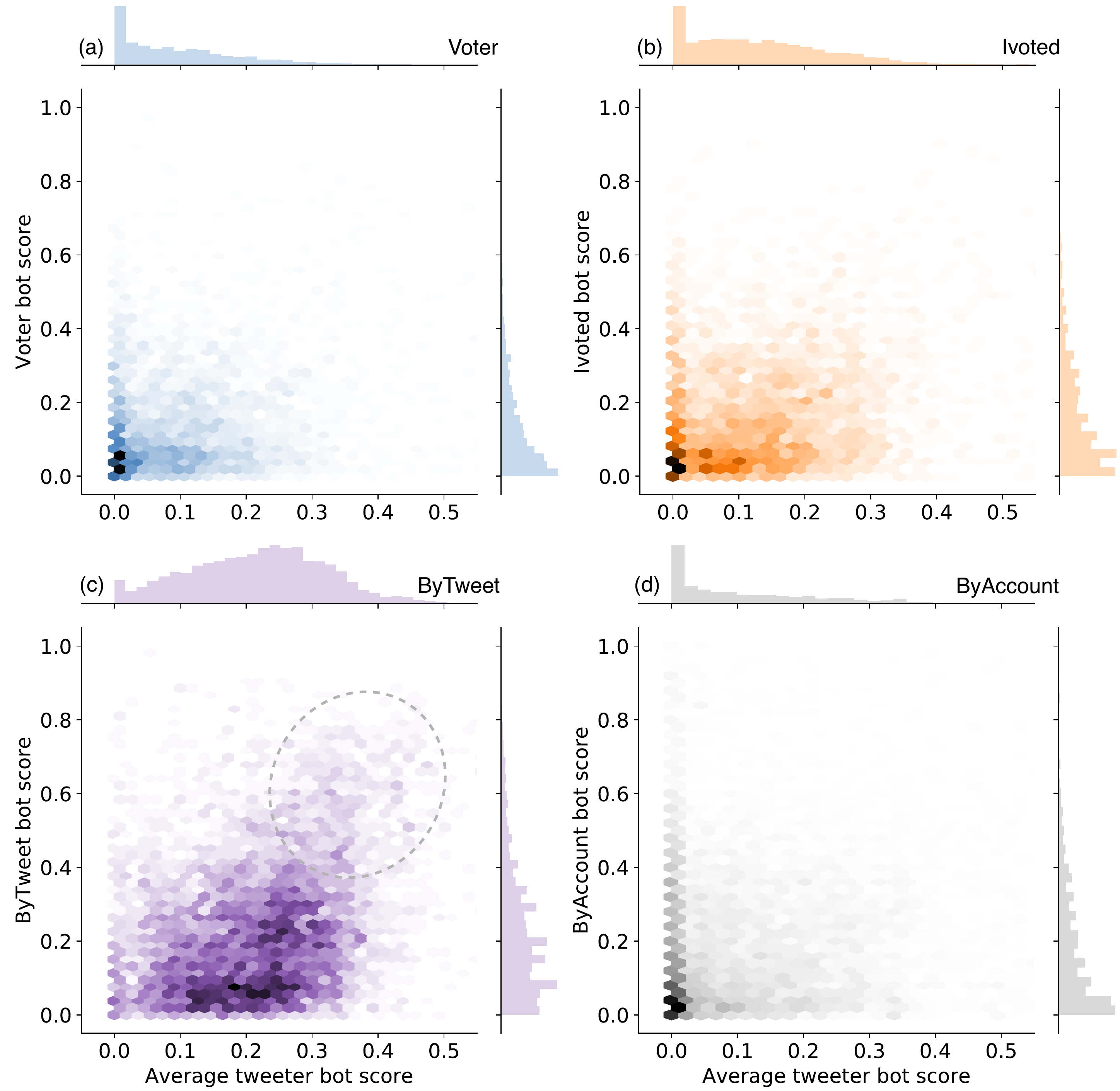}
    \caption{
    Joint distributions of the average tweeter and retweeter bot scores in (a)~\group{Voter}, (b)~\group{Ivoted}, (c)~\group{ByTweet}, and (d)~\group{ByAccount} groups.
    The plot for the \group{Voter\_auto} group is identical to the \group{Voter} plot and omitted.
    }
    \label{fig:human_bot_interaction}
\end{figure}

Since the majority of political tweets in our dataset are retweets, we also check whether the sampled accounts share messages from likely bots.
We call the accounts being retweeted ``tweeters'' and the accounts amplifying the messages ``retweeters.''
For each retweeter account in the samples, we calculate the average tweeter bot scores weighted by the number of retweets.
We plot the joint distributions of average bot scores of tweeters and bot scores of retweeters in different groups in Fig.~\ref{fig:human_bot_interaction}.

We see that the accounts amplified by the \group{ByTweet} group tend have higher average bot scores, indicating more bot-like behaviors.
On the other hand, \group{Voter}, \group{Voter\_auto}, and \group{ByAccount} groups seem to mainly engage with human-like accounts.
\group{Ivoted} lies in the middle.
We also observe increased density in the top-right region (highlighted by an oval) in Fig.~\ref{fig:human_bot_interaction}(c), suggesting interactions among bot-like accounts in the \group{ByTweet} group.

\subsection{Abnormal Tweet Deletion}

\begin{figure}
    \centering
    \includegraphics[width=\columnwidth]{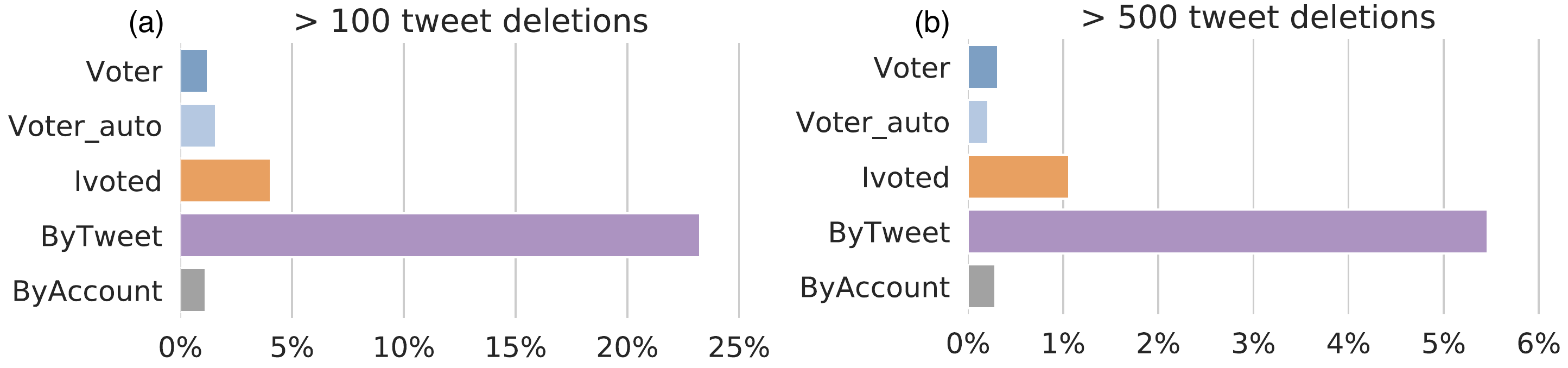}
    \caption{
    Percentages of accounts that delete more than (a)~100 and (b)~500 tweets at once in different groups.
    }
    \label{fig:tweet_deletion}
\end{figure}

Twitter users have the right to remove their content and may do so for various legitimate reasons, from fixing typos to removing regrettable content \cite{almuhimedi2013tweets,zhou2016tweet}.
However, this feature can be abused. 
For example, some politicians delete old tweets for public image management \cite{meeks2018tweeted}. Trolls and bots may systematically delete their tweets to make their actions and intentions harder to detect \cite{zannettou2019disinformation,yang2019arming}.

Since we didn't explicitly collect deletion events, we infer them from our data. User objects that come with the tweets are snapshots of the account profiles at the time of tweet collection. A decrease in the total number of status counts between two consecutive tweets in the collection indicates a tweet deletion event. Note that not all deletion events can be detected with this method. We record the maximum drop in status counts for each account.

To analyze abnormal tweet deletion behaviors, we first need to establish a reference deletion frequency. We adopt numbers from a previous study showing that Twitter users delete 7 to 11 tweets on average in a week \cite{almuhimedi2013tweets}. 
In our case, the data collection continued for about three months, so we consider a status count drop of 100 or more tweets as a conservative threshold of abnormal behavior.
Fig.~\ref{fig:tweet_deletion} shows the percentage of accounts with abnormal tweet deletion behaviors in each group. 
The \group{ByTweet} group has a much higher ratio of accounts that exhibit abnormal tweet deletion behaviors; this result is robust with respect to higher thresholds. 
%Among the remaining groups, \group{Ivoted} has a higher rate than others.

\subsection{Accounts Suspension}

\begin{figure}
    \centering
    \includegraphics[width=0.65\columnwidth]{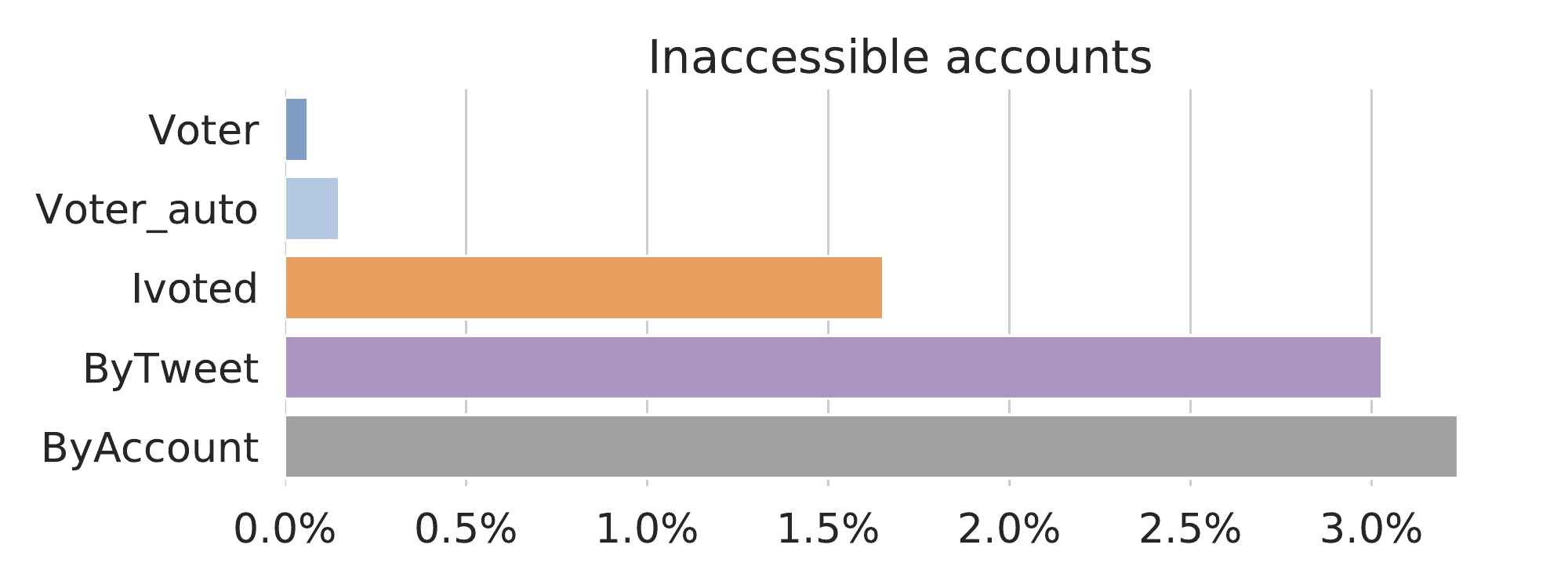}
    \caption{
    Percentages of accounts that were inaccessible as of Jan 5, 2019 in different groups. 
    }
    \label{fig:suspension_rate}
\end{figure}

In the 2018 midterm election season, Twitter reinforced their actions against suspicious activities and suspended many accounts that violated their policy.\footnote{\url{https://blog.twitter.com/en_us/topics/company/2019/18_midterm_review.html}}
In line with previous studies \cite{alothali2018detecting}, we consider being suspended by the platform as a signal of illicit behaviors.  
We checked the status of the accounts on Jan 5, 2019 through the Twitter user lookup API and plot the percentages of inaccessible accounts in different groups in Fig.~\ref{fig:suspension_rate}.
Being inaccessible means the target account is either suspended, deleted, or protected.
Users can delete or protect their accounts, so inaccessible accounts are not necessarily suspicious. 
However, such actions are very rare compared to suspension, so we use inaccessibility as a proxy for suspension.

Interestingly, \group{ByAccount} has the highest suspension rate this time followed closely by \group{ByTweet}.
This suggests that from Twitter's policy standpoint, hyperactivity is neither a flag of illicit behavior nor correlated with illicit behaviors. 
Accounts in \group{Ivoted} also suffer from suspensions to some extent, while most accounts in \group{Voter} and \group{Voter\_auto} are not affected.

\subsection{Sharing Low-Credibility Information}

Concern about low-credibility information spreading on social media has been growing ever since the 2016 U.S. presidential election.
It is therefore interesting to evaluate the level of involvement for accounts in different groups.
We focus on the URLs extracted from the collected tweets and evaluate the credibility of the content at the domain level following the literature~\cite{shao2018spread,grinberg2019fake,guess2019less,pennycook2019fighting,bovet2019influence,vosoughi2018spread}.
We compile a list of low-credibility domains from recent research papers.
A domain is labeled as low-credibility if it is either (a)~labeled as low-credibility by \citet{shao2018spread}; (b)~labeled as either ``fakenews'' or ``hyperpartisan'' by \citet{pennycook2019fighting}; (c)~labeled as ``fakenews,'' ``extremeright,'' or ``extremeleft'' by \citet{bovet2019influence}; or (d)~labeled as ``Black,'' ``Red,'' or ``Satire'' by \citet{grinberg2019fake}.  
This procedure yields 570 low-credibility domains.
Note that we treat hyperpartisan domains as low-credibility.

We follow the same procedure as in the political alignment estimation to obtain the URL domains from each tweet.
We exclude from the analysis URLs linking to Twitter itself and other social media like Facebook and YouTube.
Most other domains are news outlets. 
With the help of the list, we are able to characterize the sharing of low-credibility information by the accounts.

\begin{figure}
    \centering
    \includegraphics[width=\columnwidth]{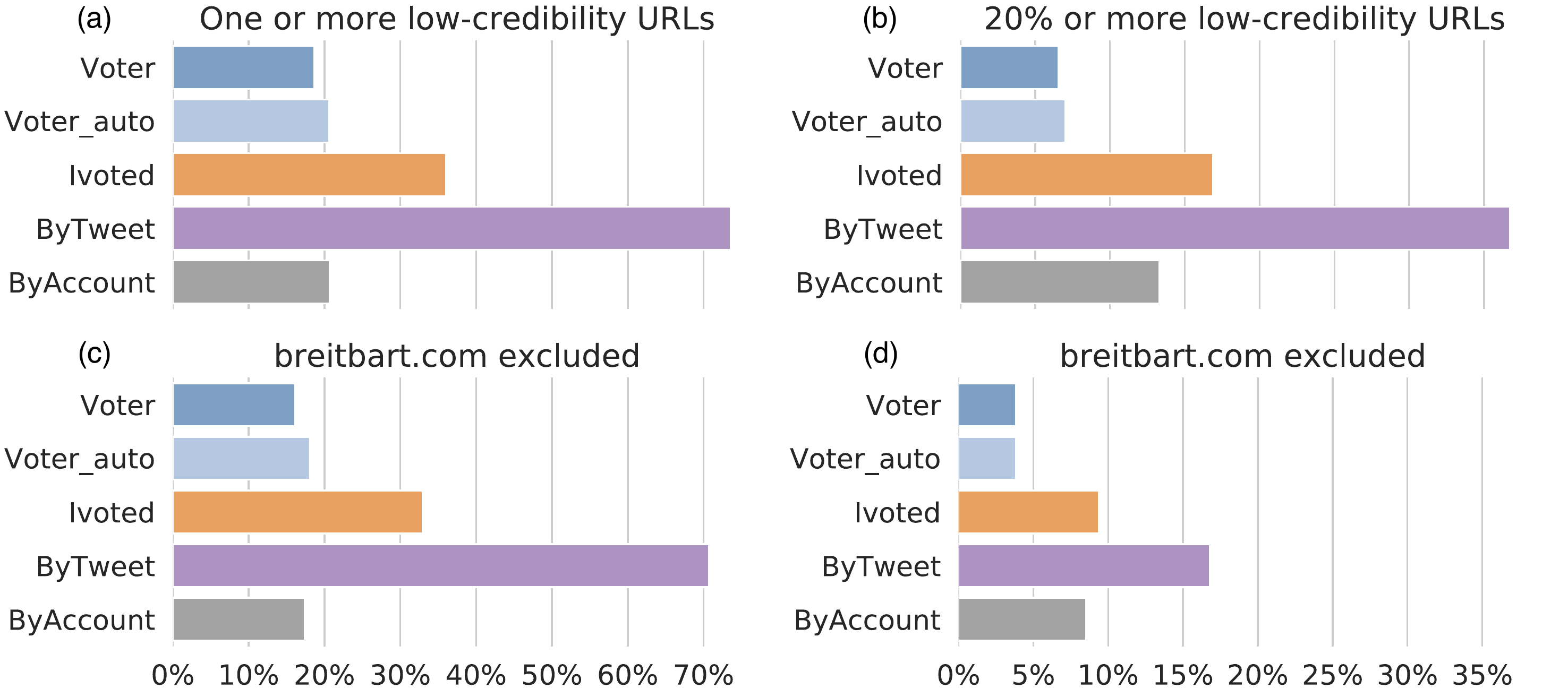}
    \caption{
    (a)~Percentages of accounts in different groups sharing at least one URL from low-credibility sources.
    (b)~Percentages of accounts in different groups having at least 20\% of shared URLs from low-credibility sources.
    (c)~Same as (a) but breitbart.com is excluded.
    (d)~Same as (b) but breitbart.com is excluded.
    }
    \label{fig:low_credibility_share}
\end{figure}

First, we calculate the percentage of accounts that shared at least one low-credibility URL for each group and plot the results in Fig.~\ref{fig:low_credibility_share}(a).
We find that the number for \group{ByTweet} is alarmingly high---over 70\% of the accounts had shared links to low-credibility sources.
The other groups have lower percentages.
We notice that breitbart.com appears much more frequently than other low-credibility domains in our dataset.
To make sure the pattern observed in Fig.~\ref{fig:low_credibility_share}(a) is not dominated by one hyperpartisan source, we perform a robustness test by excluding this domain and show the results in Fig.~\ref{fig:low_credibility_share}(c).
The proportions decrease slightly for all groups, but the results are unaffected. 

The higher portion of links to low-credibility sources shared by the \group{ByTweet} group could be trivially due to the higher number of tweets produced by accounts in this group. 
To exclude this possibility, let us also consider the proportion of low-credibility URLs shared by each account. 
In Fig.~\ref{fig:low_credibility_share}(b), we show the percentages of accounts with more than 20\% of their shared links from low-credibility sources. The results are qualitatively similar, but the percentage of accounts reduces by about half. 
We also perform the same robustness check by excluding breitbart.com and plot the results in Fig.~\ref{fig:low_credibility_share}(d).
The percentages drop further, but the general pattern is similar. 

\begin{figure}
    \centering
    \includegraphics[width=\columnwidth]{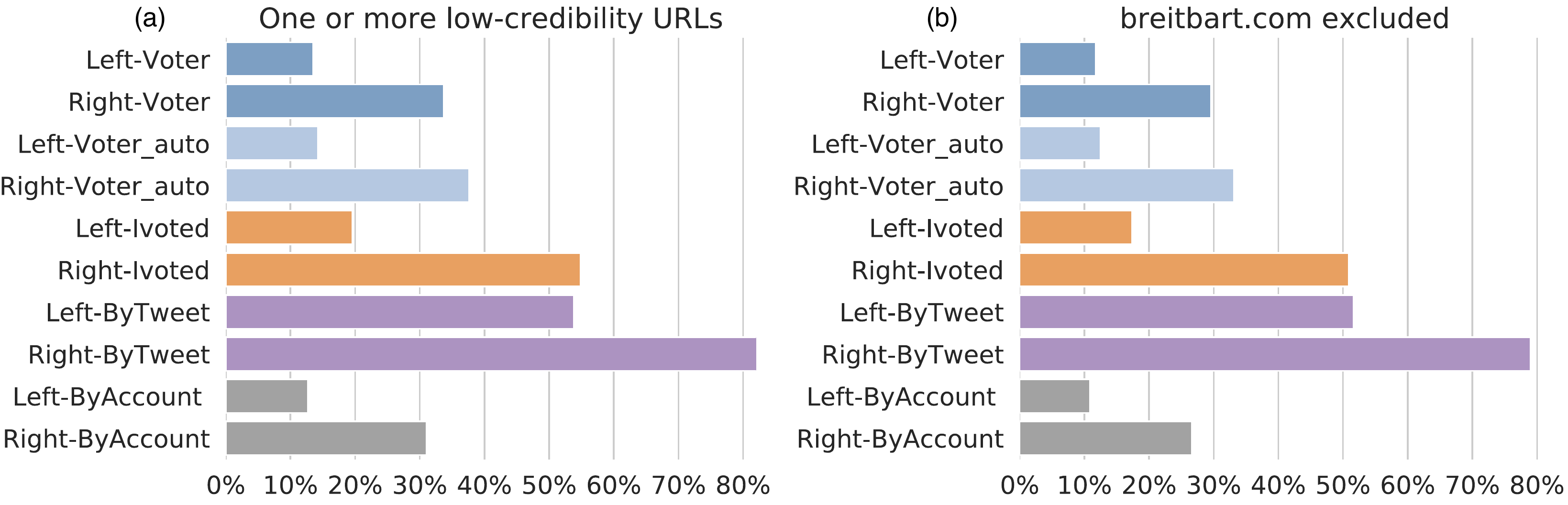}
    \caption{
    (a)~Percentages of accounts in different sub-groups that shared at least one URL from low-credibility sources.
    (b)~Same as (a) but breitbart.com is excluded.
    }
    \label{fig:political_low_credibility}
\end{figure}

In addition to treating each group as a whole, we can break them down by political alignment. 
We report the results for left- and right-leaning accounts from each group in Fig.~\ref{fig:political_low_credibility}.
By comparing accounts with the same political alignment across the groups, we find that the same patterns as before: \group{ByTweet} accounts have the highest rate of low-credibility content.
Within each group, the right-leaning accounts are more likely to share links to low-credibility sources. 
We perform the same analysis using the proportion of low-credibility URLs (not shown), with identical observations. 

\section{Discussion}

The present findings highlight the pitfalls of sampling biases when analyzing social media actors in the context of political elections. The Twitter data stream is dominated by hyperactive accounts, who introduce political bias and inauthentic signals. 

As hyperactive accounts flood the network with new content and aggressively expand their social networks, normal Twitter users can be disproportionately influenced by them. This can happen via direct exposure, indirectly via social engagement metrics, or through platform mechanisms like trending topics and feed ranking algorithms, which use social engagement metrics as signals.

The picture of the political discussion that one obtains from Twitter is subject to distortion and manipulation by inauthentic actors. This is shown by the higher levels of automated behaviors and deleted tweets observed when we naively sample content from the Twitter data stream. 

Sampling bias due to hyperactive accounts leads to an overestimation of the prevalence of low-credibility content. Joint with our finding that hyperactive accounts are disproportionately conservative, this result suggests that questionable content is preferentially shared by conservative users, in line with prior findings~\cite{grinberg2019fake,Nyhan2020}. 

To probe genuine opinions from just legitimate users, proper filtering of Twitter accounts is necessary. We explore several options.
The ideal solution is to identify real-world voters on Twitter and focus on them.
In this paper, we propose to use the \hashtag{ivoted} hashtag and corresponding pictures to find self-identified voters.
Our analyses confirm that manually identified voters are least likely to display suspicious behaviors.
We also evaluate the performance of two more efficient variants of this method. 
Using just the \hashtag{ivoted} hashtag \cite{deb2019perils} is simplest and cheapest, but results based on this sampling method are still subject to some bias along all the dimensions studied. 
The automated approach is also efficient in that no human resources are required for manual annotation, at the cost of losing some precision and recall. Our systematic comparison reveals that this sampling method is as reliable as the manual method: the \group{Voter\_auto} group is almost identical to \group{Voter} along all of the dimensions studied.

Our proposed manual and automated voter identification methods are efficient and inexpensive compared to traditional approaches. 
They can be applied to future elections in the U.S. and other countries with similar social norms around voting. 
The methods can be applied to other platforms like Instagram and Facebook where people share photos of themselves wearing ``I Voted'' stickers.

Despite the above advantages, our methods have some limitations.
First, unlike survey-based methods, our approach cannot provide accurate demographic information about the voters.
Second, only voters who are willing to reveal their voting actions online can be identified. This could introduce biases based on, say, age and privacy preferences.
Finally, the proposed methods could be exploited by malicious actors, e.g., to target voter suppression efforts.

Our analyses of the \group{ByAccount} group suggest that randomly sampling accounts with equal weights could be another effective way to minimize bias and inauthentic signals. 
Accounts in the \group{ByAccount} group are similar to likely voters and only exhibit mild suspicious behaviors, although they do have a high suspension rate.
For high quality data, one may consider applying bot and troll detection algorithms in the literature to identify and remove likely inauthentic accounts from such a sample \cite{yang2019arming,addawood2019linguistic}.

\section{Conclusion}

In this paper, we propose a method to identify voters on Twitter and systemically compare their behaviors with different samples of Twitter accounts in the context of U.S. midterm elections.
We find that compared to likely voters and randomly selected accounts, accounts sampled from the Twitter stream tend to be hyperactive and markedly more conservative. They also exhibit various suspicious behaviors, are involved with more automated activities, and share more links to low-credibility sources. 
Researchers and journalists alike who analyze political issues using social media data must be mindful of these pitfalls of sampling biases. 

%%
%% The acknowledgments section is defined using the "acks" environment
%% (and NOT an unnumbered section). This ensures the proper
%% identification of the section in the article metadata, and the
%% consistent spelling of the heading.

\begin{acks}
The authors thank Y.-Y. Ahn for useful comments; D. Pacheco and W. Chen for assistance.
%% FUNDING?
\end{acks}

%%
%% The next two lines define the bibliography style to be used, and
%% the bibliography file.
\bibliographystyle{ACM-Reference-Format}
\bibliography{ref}

%%
%% If your work has an appendix, this is the place to put it.
%\appendix

%\section{Research Methods}

\end{document}